\documentclass[aps,prd,floatfix,twocolumn,superscriptaddress,nofootinbib,a4paper]{revtex4-1}
\pdfoutput=1

\usepackage{amsmath}
\usepackage{amssymb}
\usepackage{graphicx}
\usepackage{hyperref}
\usepackage{color}
\usepackage{xcolor}
\usepackage[T1]{fontenc} 
\usepackage{slashed}

\usepackage[normalem]{ulem} 

\definecolor{blus}{cmyk}{1,1,0,0.6}
\hypersetup{colorlinks,bookmarksopen,bookmarksnumbered,linkcolor=blus,pdfstartview=FitH,urlcolor=blus,citecolor=blus}

\allowdisplaybreaks

\newcommand{\AddrOXF}{%
Rudolf Peierls Centre for Theoretical Physics, University of Oxford, Parks Road, Oxford OX1 3PU, UK
}

\begin{document}

\title{New bounds and future prospects for axion force searches at Penning trap experiments}

\author{Xing Fan}
\email{xing.fan@northwestern.edu}
 \affiliation{Center for Fundamental Physics, Department of Physics and Astronomy, Northwestern University, Evanston, Illinois 60208, USA}

\author{Mario Reig}\email{mario.reiglopez@physics.ox.ac.uk}
\affiliation{\AddrOXF}

\begin{abstract}
\noindent  In this note we consider Penning trap experiments as probes of axion-mediated forces. We show that the current measurement of electron's $g$-factor already sets a new exclusion limit for monopole-dipole axion forces acting on the electron spin. We also show that the Penning trap's capability of switching an electron and a positron can isolate the effect of an axion force and suppress systematic effects.
\end{abstract}

\maketitle

%

Axions are well-motivated new physics candidates that are very feebly coupled with the Standard Model (SM).
These light particles can generate small but macroscopic effects such as long-range forces \cite{Moody:1984ba}. While a Coulombic force is absent in the limit of $CP$-preserving interactions, a spin-dependent force can be produced from the derivative coupling with fermions. Axions therefore mediate a dipole-dipole interaction between polarised matter. In the presence of small $CP$-violating interactions,  axions can also mediate a monopole-dipole force with interesting phenomenological consequences, while monopole-monopole interactions remain very suppressed. 

In this note we consider axion models with the following interactions:
\begin{equation}\label{eq:lag}
    \mathcal{L}\supset g^N_s \phi \Bar{N}N+c_\psi\frac{\partial_\mu \phi}{f_\phi}\Bar{\psi}\gamma^\mu\gamma^5\psi\,,
\end{equation}
with $f_\phi$ being the axion decay constant, and $c_\psi$ a coupling constant. The couplings $g^N_s$ and $g_p^\psi=c_\psi\frac{m_\psi}{f_\phi}$ correspond to the the $CP$-violating scalar coupling to nucleons, $N$, and the $CP$-conserving dipole coupling to the fermion $\psi$, and will be used on the following. In particular, we will be interested in the effects of an axion gradient coupled to the spin of electrons and positrons.

Through the first term, $g^N_s\phi \bar{N}N$, nucleons source a coherent axion field\footnote{In the case of the QCD axion, the impact on this coupling of BSM sources of CP violation has been recently studied in detail in \cite{Dekens:2022gha,Plakkot:2023pui}. In this case, for the QCD axion, EDM searches provide the most stringent constraints. In this paper, however, we will consider generic axion-like particles for which the neutron EDM constrain does not necessarily apply.}. In the non-relativistic limit, through the second term in Eq.(\ref{eq:lag}), such coherent field gives rise to the following energy shift (see \cite{Graham:2017ivz} for a review):
\begin{equation}\label{eq:monopole-dipole_Hamiltonian}
    H_{\phi}=-\frac{g_p^\psi}{m_\psi}\boldsymbol{\nabla} \phi \cdot \mathbf{S}\,.
\end{equation}

This kind of interactions mediated by axions, known as monopole-dipole forces, are usually given in terms of the non-relativistic potential:
\begin{align}
    V(r)
    &=
    \frac{g^N_s g^\psi_p}{8\pi m_\psi}
    \left(
    \frac{1}{\lambda_\phi r}+\frac{1}{r^2}
    \right)
    e^{-m_\phi r}\mathbf{S}\cdot \hat{r}\,.
\end{align}
The mass of the axion sets the effective reach of the force $\lambda_\phi\sim m_\phi^{-1}$, above which the potential is exponentially suppressed. 

The presence of a sourced axion background therefore generates a spin evolution characterised by the precession frequency
\begin{equation}
    \boldsymbol{\omega}_\text{axion}=-\frac{g_p^\psi}{m_\psi}\boldsymbol{\nabla}\phi\,,
\end{equation}
which implies that the spin will precess around the direction of the gradient $\boldsymbol{\nabla} \phi$. 
Here, we define the angular momentum as a vector.
This axion-induced angular frequency $\boldsymbol{\omega}_\text{axion}$ does not depend on whether one considers particles or anti-particles.
The origin of this gradient can be either a dense material acting as a source test mass, or a radially oriented gradient coming from the center of the earth. 
Using movable test masses at a fixed frequency is an interesting way to reject systematic effects which has been used and proposed as an experimental scheme \cite{Lee:2018vaq,Crescini:2020ykp,Arvanitaki:2014dfa}.
Searching for static signals such as the earth gradient benefits from a much larger source mass, but is a more complicated task, requiring a better rejection of background and systematic effects.
Precision experiments such as atomic or molecular beams \cite{Agrawal:2023lmw}, and storage rings\footnote{Storage rings are also appropriate to search for axion DM \cite{Graham:2020kai}. See \cite{JEDI:2022hxa} for a recent search. In this note however, we focus on axion forces which do not require any cosmological abundance of axions, nor any kind of frequency scan.} \cite{axionSearchStorageRing2023} are appropriate to that end. Constraining monopole-dipole interactions from purely astrophysical observations has also been considered recently in \cite{Poddar:2023bgk}.
In this note, we show that Penning traps particularly provide a unique way to probe the axion gradient because of their capability to switch particle and antiparticle rapidly.

In a Penning trap, a magnetic field is applied in the $z$ direction $\mathbf{B}=B\hat{z}$, and the $g$-factor is measured from the ratio of the spin-precession frequency $\boldsymbol{\omega}_s=-\frac{g}{2}\frac{qB}{m}\hat{z}$ and the cyclotron frequency $\boldsymbol{\omega}_c=-\frac{qB}{m}\hat{z}$ ($q$ is the trapped particle's charge and $m$ is its mass), also defined as vectors.
We take the electron-positron trap as an example, but the argument below applies to proton-antiproton and other particles as well.
The charge is $q=-e$ ($+e$) for an electron (positron), so $\boldsymbol{\omega}_c$ and $\boldsymbol{\omega}_s$ point at $\hat{z}$ ($-\hat{z}$).
The anomaly frequency $\boldsymbol{\omega}_a=\boldsymbol{\omega}_s-\boldsymbol{\omega}_c$ is experimentally measured, and the $g$-factor is obtained from the relation
\begin{equation}
    \frac{g}{2}=\frac{|\boldsymbol{\omega}_s|}{|\boldsymbol{\omega}_c|}=1+\frac{|\boldsymbol{\omega}_a|}{|\boldsymbol{\omega}_c|}.
\end{equation}

\begin{figure*}[t]
	\centering
	\includegraphics[width=0.7\textwidth]{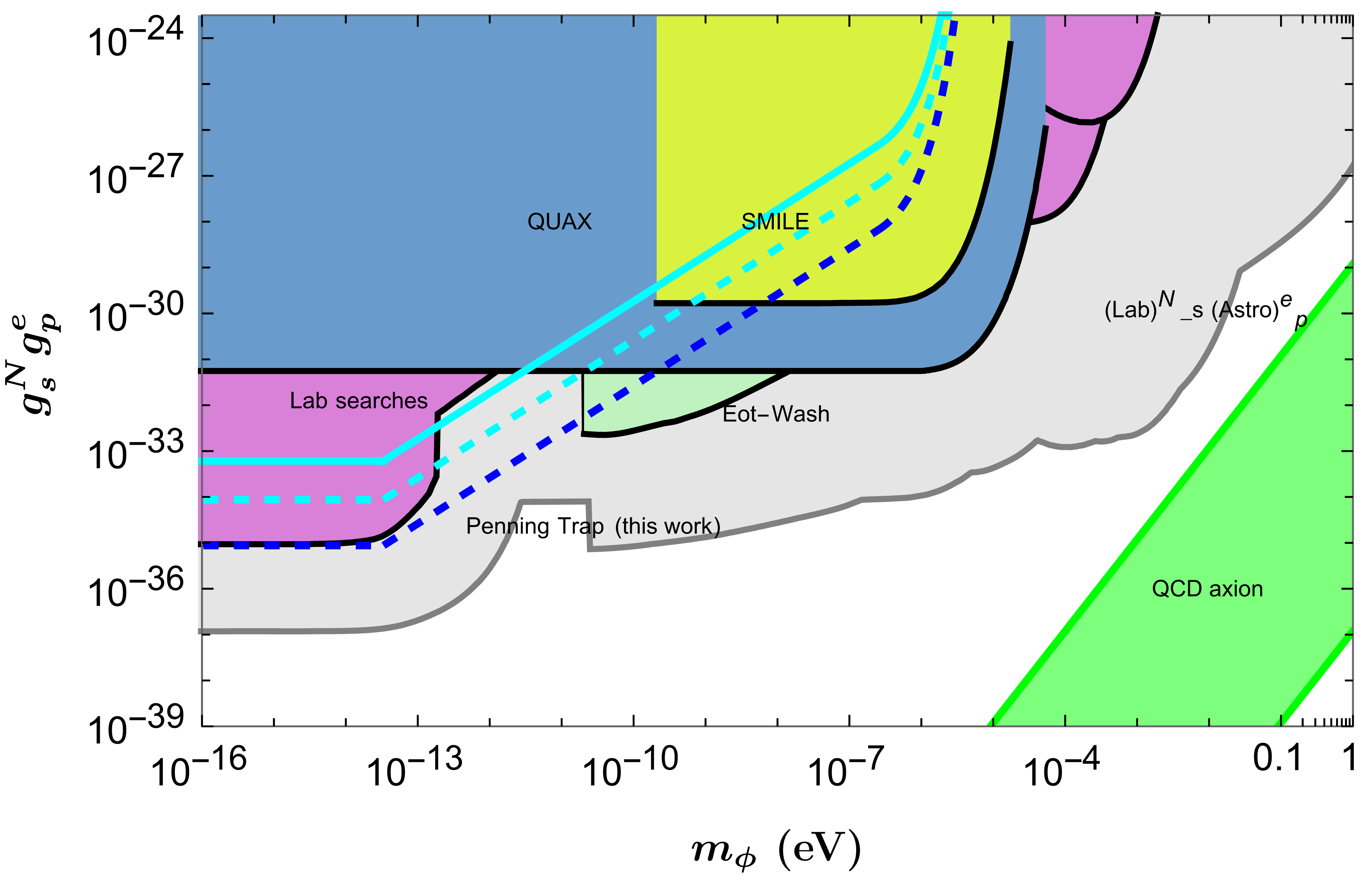}
		\caption{
  Axion-mediated monopole-dipole forces on electrons at different spin precession experiments, including \cite{Crescini:2020ykp,Lee:2018vaq,Heckel:2008hw,PhysRevLett.67.1735,PhysRevLett.106.041801,PhysRevLett.115.201801}. The QCD axion band is shown in light green, taking $\theta_{\text{eff}}$ to lie in the range $10^{-20}<\theta_{\text{eff}}<10^{-10}$. Cyan corresponds to the current sensitivity at the Penning trap experiment \cite{FanElectronMagneticMoment2023}. The dashed cyan will correspond to the sensitivity when the SM discrepancy is solved. Dark blue (dashed) corresponds to future upgrades including the use of positrons.
 See \cite{OHare:2020wah} for details on the different experimental schemes.}
	\label{fig:axion_electron_forces}
\end{figure*}

In the presence of Eq.(\ref{eq:lag}), the axion field will modify the measured $g$-factor as
\begin{equation}
    \frac{g}{2}=1+\frac{|\boldsymbol{\omega}_a+\boldsymbol{\omega}_\text{axion}|}{|\boldsymbol{\omega}_c|}.
    \label{eq:AxionFrequencyShift}
\end{equation}
Since the $g$-factor has a very precise Standard Model's (SM) prediction \cite{atomsTheoryReview2019,QED_C10_nio,MuellerAlpha2018,RbAlpha2020}, we can use the measured electron's $g$-factor to set a limit on axion with the existing data.
In the recent electron's $g$-factor measurement \cite{FanElectronMagneticMoment2023}, the magnetic field axis $\hat{z}$ is aligned with the earth's radial direction ($\hat{z}\parallel\boldsymbol{\omega}_\text{axion}$), and the anomaly frequency is measured at $|\delta\boldsymbol{\omega}_a|=0.1$~rad/s resolution.
The result is consistent with the SM's prediction \cite{atomsTheoryReview2019} within $|\delta\boldsymbol{\omega}_a|=0.7$~rad/s, which is limited by the discrepancy of the fine-structure measurements \cite{MuellerAlpha2018,RbAlpha2020}.
The consistent result sets a new limit for the axion-induced coupling $g^{N}_sg^{e}_p$, shown by the solid cyan line in Fig.~\ref{fig:axion_electron_forces}.
The possible limit if the fine-structure constant discrepancy is resolved ($|\delta\boldsymbol{\omega}_a|=0.1$~rad/s) is also shown with a dashed cyan line.

An elegant way to evade the SM prediction's uncertainty is to compare the $g$-factors of an electron and a positron in the same Penning trap\footnote{Note that since the magnetic field is maintained during the switch, the electron and the positron have opposite rotational directions.}. 
If $\hat{z}$ is aligned with the earth gravity's direction, Eq.~\eqref{eq:AxionFrequencyShift} transforms as
\begin{eqnarray} 
    |\boldsymbol{\omega}_a+\boldsymbol{\omega}_\text{axion}|&=&\left|\pm\frac{g-2}{2}\frac{eB}{m}\hat{z}+\boldsymbol{\omega}_\text{axion}\right|\nonumber\\
    &=&\frac{g-2}{2}\frac{eB}{m}\pm|\boldsymbol{\omega}_\text{axion}|,
\end{eqnarray}
where the $\pm$ sign corresponds to electron and positron respectively, and we used that $\frac{g-2}{2}\frac{eB}{m}\gg|\boldsymbol{\omega}_\text{axion}|$.
Taking the difference of the $g$-factors of an electron and a positron in the same Penning trap directly extracts the effect of $|\boldsymbol{\omega}_\text{axion}|$.
We estimate that one can achieve $|\delta\boldsymbol{\omega}_a|<0.01$~rad/s (dark blue line in Fig.~\ref{fig:axion_electron_forces}) in this scheme and describe the details below.

A stacked trap with an electron-park trap, a $g$-factor measurement trap, and a positron-park trap is suitable for electron and positron's $g$-factor comparison ~\cite{BASEDemonstrationDoublePenningTrapTechnique,BASEProtonMagneticMoment2017,BASEPBarMagneticMoment2017,BaseChargeToMassRatio2022,PENTATRAPProposal2012,ProtonAtomicMassFabiran2017,DeutronHD+Ion2020}.
When the $g$-factor of an electron (positron) is measured, the electron (positron) is transferred to the center measurement trap, and the positron (electron) is stored in its park trap.
Importantly, $g$-factors of both particles are measured in the same trap, but with an opposite trap voltage polarity.
The switching of an electron and a positron is done by changing the electrode voltages (without a change of the magnetic field), typically on a minute time-scale.
This switching has been already used in the proton-antiproton Penning traps, but unfortunately, the most precise proton and antiproton traps are located horizontally ($\hat{z}\perp\boldsymbol{\omega}_\text{axion}$), so they are not sensitive to the axion field from the earth as it is \cite{BASEProtonMagneticMoment2017, BASEPBarMagneticMoment2017}.

The largest systematic shift in the electron and positron's $g$-factor measurements is also canceled---the microwave cavity shift or image charge shift \cite{RenormalizedModesPRL,RenormalizedModesPRA}.
It originates from the image charge on the electrode by the trapped particle, which generates an electric field in the trap and changes its own cyclotron frequency \cite{ImageChargeShiftUW2006,ImageChargeShiftDiscovery1989,ImageChargeShiftCylindricalCalculation2001,SiGFactorMeasurement2013,ImageChargeShiftSven2019}.
The shift depends on the square of the trapped ion's charge $q^2$, so is the same for an electron and a positron if the measurement is done at the same location.
A non-reversing voltage will shift the exact locations of the electron and positron and cause an imperfect cancellation.
However, taking the geometry in the electron's $g$-factor measurement \cite{FanElectronMagneticMoment2023}, even a 100~mV of non-reversing voltage shifts the trapped particle's location only by 5~$\mu$m.
The corresponding imperfect cancellation is estimated to be less than $|\delta\boldsymbol{\omega}_a|<0.005$~rad/s. 

With the achieved statistical sensitivity in the recent measurement, $|\delta\boldsymbol\omega_a|=0.16$~rad/s per day, the target precision $|\delta\boldsymbol\omega_a|<0.01$~rad/s can be reached with one year of data-taking.
The development of a more harmonic trap and a SQUID detector will improve the sensitivity further \cite{ThesisFan}. 
Using the same analysis method for both electron and positron could also eliminate the fitting-model dependent systematic shift.
The long-term magnetic field drift is canceled by measuring both spin frequency and cyclotron frequency in the measurement trap.

In conclusion, we point out that the measured electron's $g$-factor can already set a limit on the axion-induced coupling.
The precision is limited by the discrepancy of the fine-structure measurements.
However, a comparison of an electron and a positron's $g$-factors in the same trap will evade the limit from the fine-structure constant.
This particle-antiparticle switch is a unique feature of Penning traps, and also removes dominant image charge systematic shift in the $g$-factor measurement. 
The comparison of an electron and a positron's anomaly frequency at $0.01$~rad/s will further improve the reach of the Penning trap experiment to axion-mediated monopole-dipole forces by an order of magnitude.

\section*{Acknowledgments}
MR would like to thank Prateek Agrawal, Nick Hutzler, David Kaplan, On Kim, and Surjeet Rajendran for an amazing collaboration on related projects. This article is based upon work from COST Action COSMIC WISPers CA21106, 
supported by COST (European Cooperation in Science and Technology), U.S. DOE, Office of Science, National QIS Research Centers, Superconducting Quantum Materials and Systems Center (SQMS), NSF Grants No.~PHY-2110565, and the John Templeton Foundation Grants No.~61906 and No.~61039. 
%
%

\noindent 

\bibliography{newrefs_axion.bib}

\end{document}